\makeatletter
\@ifundefined{@parse@version@dash}{%
\def\@parse@version#1{\@parse@version@0#1}
\def\@parse@version@#1/#2/#3#4#5\@nil{%
\@parse@version@dash#1-#2-#3#4\@nil}
\def\@parse@version@dash#1-#2-#3#4#5\@nil{%
  \if\relax#2\relax\else#1\fi#2#3#4 }
}{}
\makeatother
\documentclass[twocolumn,secnumarabic,amsmath,amssymb,aps,prr,superscriptaddress,eqsecnum,graphicx,dvipdfmx]{revtex4-2}

\usepackage{graphicx}
\usepackage{bm}
\usepackage{color}
\usepackage{ulem}
\usepackage[dvipdfmx,
colorlinks=true,
urlcolor=blue,
citecolor=blue,
linkcolor=blue,
hyperfootnotes=false]{hyperref}

\begin{document}
\title{Automatic structural optimization of tree tensor networks}

\author{Toshiya Hikihara}
\affiliation{Graduate School of Science and Technology, Gunma University, Kiryu, Gunma 376-8515, Japan}
\author{Hiroshi Ueda}
\affiliation{Center for Quantum Information and Quantum Biology, Osaka University, Toyonaka 560-0043, Japan}
\affiliation{JST, PRESTO, Kawaguchi 332-0012, Japan}
\affiliation{Computational Materials Science Research Team, RIKEN Center for Computational Science (R-CCS), Kobe 650-0047, Japan}
\author{Kouichi Okunishi}
\affiliation{Department of Physics, Niigata University, Niigata 950-2181, Japan}
\author{Kenji Harada}
\affiliation{Graduate School of Informatics, Kyoto University, Kyoto 606-8501, Japan}
\author{Tomotoshi Nishino}
\affiliation{Department of Physics, Graduate School of Science, Kobe University, Kobe 657-8501, Japan}

\begin{abstract}
Tree tensor network (TTN) provides an essential theoretical framework for the practical simulation of quantum many-body systems, where the network structure defined by the connectivity of the isometry tensors plays a crucial role in improving its approximation accuracy.
In this paper, we propose a TTN algorithm that enables us to automatically optimize the network structure by local reconnections of isometries to suppress the bipartite entanglement entropy on their legs.
The algorithm can be seamlessly implemented to such a conventional TTN approach as density-matrix renormalization group.
We apply the algorithm to the inhomogeneous antiferromagnetic Heisenberg spin chain having a hierarchical spatial distribution of the interactions.
We then demonstrate that the entanglement structure embedded in the ground-state of the system can be efficiently visualized as a perfect binary tree in the optimized TTN.
%We then demonstrate that the algorithm efficiently visualizes the entanglement structure embedded in the ground state of the system in the optimized TTN structure of a perfect binary tree.
Possible improvements and applications of the algorithm are also discussed.
\end{abstract}

\date{\today}

\maketitle

\section{Introduction}\label{sec:Intro}

Tensor networks have been attracting growing interest in a variety of research fields, including condensed-matter physics, quantum information, quantum cosmology, and data science\cite{OkunishiNU2022,Ashley_2018,Orus2019,Stoudenmire_2016}. 
Under the context of quantum many-body physics, tensor networks have provided flexible representations for low-energy 
states in correlated systems, and have been employed in various theoretical approaches\cite{Cirac_2021}. 
A tensor network framework had been used earlier in the field of statistical physics\cite{Baxter1968}, but recent rapid developments in this field chiefly originated from the density-matrix renormalization group (DMRG) method\cite{White1992,White1993}.
In particular, it turned out that the matrix-product state (MPS) is variationally improved through the numerical sweeping processes in the DMRG method\cite{OstlundR1995,RommerO1997}.  

A couple of fundamental frameworks of the tensor network have been established during the first decade of this century, in which generalization of the network structure plays a key role. 
For instance, the tensor-product state\cite{NishinoHOMAG2001,GendNishino2003}, equivalently, the projected entangled-pair state\cite{VerstraeteC2004,VerstraeteWPC2006}, have been applied successfully to two-dimensional (2D) quantum lattice models and 3D statistical ones.
The multi-scale entanglement renormalization ansatz (MERA) revealed the intrinsic network structure for properly describing 1D quantum critical states\cite{Vidal2007,EvenblyV2009}. 
Moreover, the tree tensor network (TTN) framework has been established as  a solid basis for practical numerical analyses in quantum chemical problems\cite{Nakatani_2013,Murg_2010,Murg_2014,Gunst_2018,Gunst_2019,Larsson2019}, equilibrium phase and ground state studies\cite{Nagaj_2008,Tagliacozzo_2009,Silvi_2010,Li_2012,Changlani_2013,Pi_2013,Milsted_2019,Macaluso_2020,Lunts_2021,Kloss_2020}, quantum many-body systems with randomness\cite{HikiharaFS1999,GoldsboroughR2014,LinKao_2017,SekiHO2020,SekiHO2021,Ferrari_2022}, and information science\cite{Dumitrescu_2017,Cheng_2019,Wall_2021,Grelier_2022,Chen_2022,Seitz_2022}. 
Also, we note that the DMRG applied to Bethe lattice systems invokes the TTN framework\cite{Otsuka_1996,Friedman_1997,Kumar_2012}.

Suppose that we have a quantum state described by a tensor network, which is a contraction of tensors. 
In general, a tensor leg that connects a pair of tensors, which is often mentioned as auxiliary bond, with a finite dimension $\chi$ can be capable of representing the entanglement entropy (EE) up to $\ln \chi$.
Thus, if $n$ auxiliary bonds cross the boundary between a subsystem and the rest of the entire network, the bipartition EE can be $n \ln \chi$ at most. 
This upper bound controls the precision of the tensor network, when it is used as an approximation of a certain quantum state. 
For practical tensor-network approaches, the network structure is thus crucial in improving the accuracy.
%it is thus important to find the {\it natural} connectivity of tensors reflecting the intrinsic structure of the entanglement contained in the target state.

For spatially uniform 1D quantum systems, the network structure of the MPS, i.e. the linear arrangement of three-leg tensors along the chain direction, seems to fit the lattice structure of their ground-state wavefunctions. 
However, there is no guarantee, in general, that such an intuitive network structure is really suitable to the accurate representation of the target state. 
Under the presence of long-range interactions or position-dependent interactions such as random couplings, it is even difficult to imagine a good network structure in advance. 
How can we resolve the situation? 
Formulations in the strong-disorder renormalization group (SDRG)\cite{Ma_1979,Dasgupta_1980} and its tensor network generalization (tSDRG)\cite{HikiharaFS1999,GoldsboroughR2014,SekiHO2020} provide us an insight into this fundamental question. 
In the tSDRG, the network structure of tensors is determined by the energy scale contained in the effective Hamiltonians, which are sequentially generated by successive real-space renormalization group (RG) transformations.
Another variant of tSDRG that utilizes the EE in the construction of the TTN has also been clarified to generate slightly different network structure compared with the above tSDRG based on the energy scale\cite{SekiHO2021}. 
Recently, Roy {\it et al.} introduced the concept of {\it emergent geometry}, that is the entanglement structure visualized from the EEs for all the possible bipartitions of the system\cite{Roy_2020,Singha_Roy_2021,Sudipto_2021,Santalla_2022}.
The relation between the structure of entanglement in a quantum state and the geometry of tensor network has also been explored\cite{EvenblyV2011,HyattGB2017}.

As a step toward the construction of the better tensor network structures, we focus on TTNs, which do not contain any loop in the diagrammatic representation. 
An important feature of the TTN is that the bipartition of any TTN is possible just by cutting an auxiliary bond between three-leg tensors, which we call isometries following the convention\cite{Vidal2007, EvenblyV2009}. 
This means that each auxiliary bond bears the EE for the corresponding bipartition.
If there is no limitation in the dimension of auxiliary bonds, every bond can carry any amount of EE so that the TTN with arbitrary tree structure can, in principle, represent any quantum state exactly.
However, in the practical situation where the bond dimension is bounded by $\chi$, a large amount of EE exceeding $\ln \chi$ causes a loss of accuracy in representing the quantum state.
%Therefore, if there is a bond at which the EE of the target state exceeds the limit of the EE represented by the $\chi$-dimentional bond, $\ln \chi$, the excess EE missed at the bond causes a loss of accuracy.
This fact leads us to a guiding principle that one should select a TTN structure in which the EE at every bond is as small as possible so that one can minimize the loss of accuracy due to the restriction of the bond dimension.
% in representing the target state.
We call a TTN that satisfies this requirement of the least-EE principle as the {\it optimal} one in the following.
We note that the preclusion of the bonds with large EE can be partially achieved by reordering of the sites\cite{Legeza_2004,Legeza_2015,Legeza_2020,Ali_2021,Legeza_2022} or the re-adjustment of the network structure\cite{Murg_2010,Murg_2014,Szalay_2015}, based on the mutual information between sites. 
Larsson introduced a scheme to reconstruct the local tree structure by monitering the cut-off bond dimension\cite{Larsson2019}.
Very recently, Li {\it et al.} examined the efficiency of the site reordering by the successive minimization of the bond EE and/or truncation error within the MPS\cite{LiRYS2022}.
An entanglement bipartition approach to construct the optimal TTN for a given, exact wavefunction of a small system has also been proposed\cite{OkunishiUN2022}.

%Moreover, the loopless structure makes it possible to represent the TTN in the {\it canonical form} with orthonormal tensors and the corresponding singular values\cite{Vidal_2006}. 
%As in the finite-system DMRG algorithm, one can then perform variational estimation of the ground state via successive improvements of isometries \cite{Vidal_2006, Gerster_2014}, where the EEs of each auxiliary bond can be easily obtained from the singular values for the corresponding bipartition of the system.
%Our basic idea is that the EE on the auxiliary bond in the TTN can be also used as a quantitative diagnosis of its network structure; if a bond has an unnecessarily large value of EEs, the network structure might not be natural.
%Such a concentration of EE can be partially precluded by reordering of the sites\cite{Legeza_2004,Legeza_2015,Legeza_2020,Ali_2021,Legeza_2022} or the re-adjustment of the network structure\cite{Murg_2010,Murg_2014,Szalay_2015}, based on the mutual information between sites. 
%In practical variational TTN approaches, it is thus important to find out a natural TTN structure, which enables us to represent the targeted state without encountering a large value of EE on the auxiliary bond. 
%Let us call a TTN that satisfies this requirement as the {\it optimal} one in the following.

In this work, we propose an iterative algorithm that generates the optimal TTN structures automatically. 
More precisely, in the DMRG-like updating step of tensors, we determine the locally optimal connectivity of the isometry pair among three possible candidates so as to reduce the EE on the bond under consideration. (See Fig.\ref{fig:process_2}.)
During sweeps over the variational TTN wavefunction, a better network structure is thereby generated successively toward the optimal one through local reconnections of auxiliary bonds.
We apply the algorithm to the inhomogeneous antiferromagnetic Heisenberg spin chain which possesses hierarchical spatial modulation of the interaction constants, and then confirm that the perfect binary tree network emerges automatically. 
Note that the sweeping process of the algorithm can be interpreted as a visualization of the entanglement structure embedded in the TTN.
%, which is confirmed in the distribution of EE for the obtained optimal TTN.

The rest of the article is organized as follows.
We introduce the basic notations in the conventional variational TTN formulation in Sec.~\ref{sec:algorithm}.
We explain the automatic optimization scheme of the TTN structure in Sec.~\ref{sec:tree_str_opt}. 
In Sec.~\ref{sec:numerics}, we apply the proposed algorithm to the trial system, and check the 
numerical validity of the algorithm. Sec.~\ref{sec:conc} is devoted to the summary and concluding remarks.

\section{Tree tensor network}\label{sec:algorithm}

\begin{figure}
\begin{center}
\includegraphics[width = 73 mm]{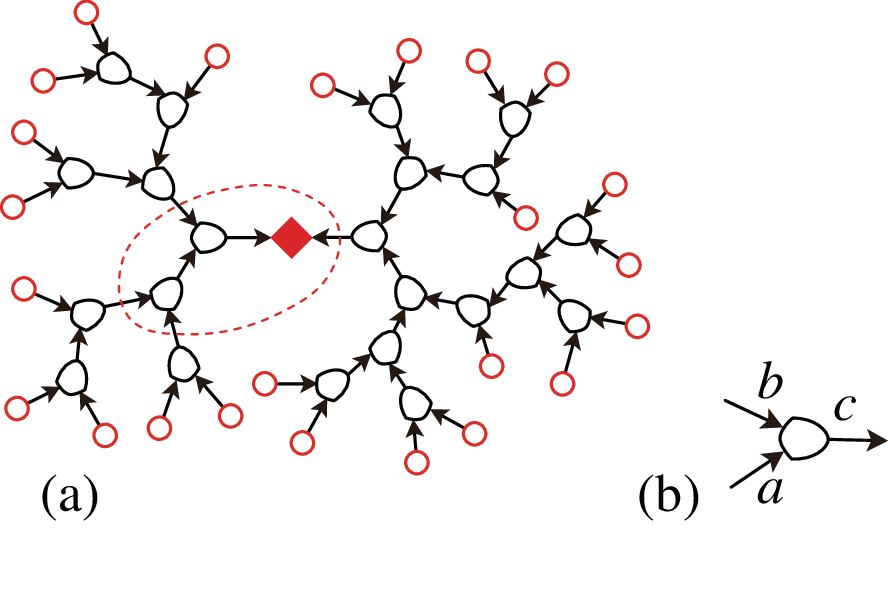}
\caption{(a) Schematic picture of the binary TTN.
The ovals (black) and circles (red) represent isometries and bare spin degrees of freedom, respectively.
The arrows are the tensor legs, which we call the (auxiliary) bonds.
The rhombus (red) represents the singular values. 
The part surrounded by the dotted curve (red) is the central area.
(b) The isometry $V_{ab}^{c}$. Directions of the bonds are shown by arrows.}
\label{fig:TTN}
\end{center}
\end{figure}

Before explaining the optimization scheme on the TTN structure, we shortly review fundamental concepts and notations in the conventional formulation of the variational TTN.  
Let us consider the binary TTN. 
Figure \ref{fig:TTN} (a) shows the schematic diagram of a TTN, which can be also viewed as the TTN representation of the ground-state wavefunction of a quantum spin model. 
In the figure, the oval and circle symbols represent isometries and bare spin degrees of freedom respectively, and they are connected to other isometries or circles through arrows representing auxiliary bonds.
As depicted in the diagram, the direction of the bond arrows flows to the {\it center} indicated by the rhombus symbol representing singular values.
As a result, the network contains no loop.
Also, we define the {\it central area} by the dotted curve surrounding the center of the network for later convenience. 

As shown in Fig.\ \ref{fig:TTN} (b), the isometry has two incoming bonds (arrows) and an outgoing one.
Let us write tensor elements of the isometry as $V_{ab}^{c}$, where $a$ and $b$ are indices for incoming bonds, and $c$ is that for the outgoing one. 
We distinguish tensors by the letters put on the bonds, if necessary. 
The isometry satisfies the orthonormal condition
\begin{eqnarray}
\sum_{ab}^{~} \bar{V}_{ab}^{c} \, V_{ab}^{c'} = \delta_{c c'}^{~} \, ,
\label{eq:orthonormal}
\end{eqnarray}
where $\bar{V}_{ab}^{c}$ is the complex conjugate of ${V}_{ab}^{c}$ and $\delta_{c c'}^{~}$ is the Kronecker delta. 
Throughout this article, we assume that the number of bond degrees of freedom, i.e. the bond dimension for $a,b$, and $c$, is bounded by a finite number $\chi$. 
For the {\it central bond} attached with the rhombus, meanwhile, we assign $k$ to specify its index for a moment. 
We then assume that singular values $D_k^{~}$ are arranged in the descending order
\begin{equation}
D_1^{~} \ge D_2^{~} \ge D_3^{~} \ge \cdots \ge D_{\chi}^{~} (\ge0)\, .
\end{equation}
Since the TTN state $| \Psi \rangle$ is written in the canonical form, the norm is expressed as
\begin{equation}
\langle \Psi | \Psi \rangle = \sum_{k = 1}^{\chi} \left( D_k^{~} \right)^2_{~} \, ,
\end{equation}
because of the orthonormal condition of the isometry\cite{Vidal_2006}. 
Here, we note that the position of the center can be moved to arbitrary bond by means of fusion processes of tensors\cite{Vidal_2006}. 

\begin{figure}
\begin{center}
\includegraphics[width = 85 mm]{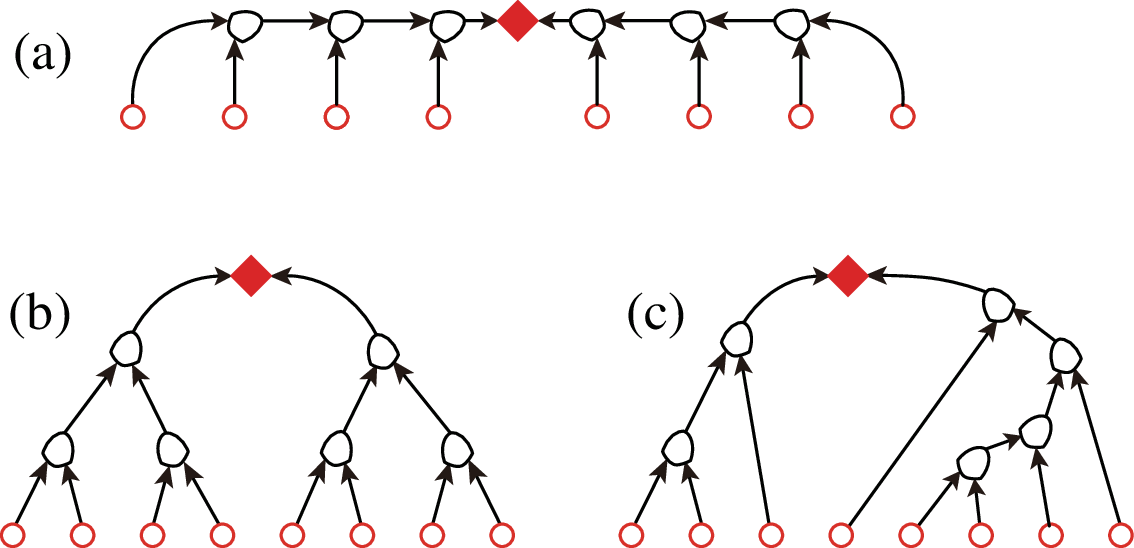}
\caption{A variety of TTN geometries. 
(a) The MPN.
(b) The perfect binary TTN.
(c) An example of the non-uniform TTN.}
\label{fig:typicalTTN}
\end{center}
\end{figure}

Given the Hamiltonian ${\hat H}$ of a system, we evaluate the variational energy,
\begin{equation}
E_{\rm var}^{~} = \frac{\langle \Psi | {\hat H} | \Psi \rangle}{\langle \Psi | \Psi \rangle} \, ,
\label{Eq_2_4}
\end{equation}
where $| \Psi \rangle$ denotes a trial quantum state in terms of TTN.
Here, the TTN representation of $| \Psi \rangle$ is not unique.
Consider TTNs for a system with eight spins as typical examples. 
Figure\ \ref{fig:typicalTTN} (a) shows the TTN of the MPS form, which is a TTN established for uniform 1D quantum systems.
We call this type of TTN the matrix-product network (MPN)\cite{note_MPN}.
Note that although the MPN has the simple chain structure,  there remains room for various ordering of the site index\cite{Legeza_2004,Legeza_2015,Legeza_2020,Ali_2021,Legeza_2022,LiRYS2022,ChanHG2002,MoritzHR2005}.
Figure\ \ref{fig:typicalTTN} (b) is the perfect binary TTN with a hierarchical structure, which is often used in the framework of real-space RGs. 
Figure\ \ref{fig:typicalTTN} (c) shows a non-uniform TTN structure, which often appears in  tSDRGs for random systems. 
These examples suggest that we have several options in choosing an appropriate TTN structure for obtaining better variational energy $E_{\rm var}^{~}$ of Eq.~(\ref{Eq_2_4}).

\begin{figure}
\begin{center}
\includegraphics[width = 75 mm]{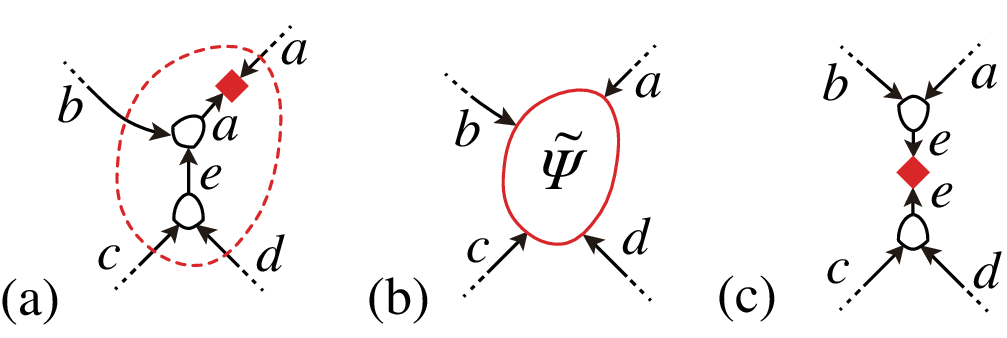}
\caption{Local improvement process. 
(a) The current central bond $a$ and two isometries connected to the new central bond $e$.
(b) The renormalized ground-state wavefunction ${\tilde \Psi}_{abcd}^{~}$.
(c) Updated isometries connected to the new central bond $e$.}
\label{fig:process_1}
\end{center}
\end{figure}

If the network structure of the TTN is fixed, the variational minimum of $E_{\rm var}^{~}$ can be obtained by the successive improvement of isometries, which is quite similar to that of the finite-system DMRG algorithm~\cite{Vidal_2006,Murg_2010,Murg_2014,Gerster_2014,Szalay_2015,Cao_2021}. 
We briefly explain the local improvement procedure of the isometries at the center of the TTN, which is also essential for constructing the local reconnection algorithm of isometries discussed in Sec.\ \ref{sec:tree_str_opt}. 
Figure\ \ref{fig:process_1} (a) shows the central area in the TTN to move the central bond from $a$ to $e$. 
At first, the singular value $D_a^{~}$ is located at the central bond $a$, where two isometries $V_{be}^{a}$ and $V_{cd}^{e}$ are connected by the bond $e$, which will be a new central bond. 
The bonds $a$, $b$, $c$, and $d$ have arrows incoming to the central area surrounded by the dotted curve, and then the isometries outside the central area define the corresponding environment. 
We can then construct the renormalized Hamiltonian ${\tilde H}$ by recursively applying the isometries in the environment to ${\hat H}$ from the boundary of the TTN toward the center. 
In practical computations, this RG transformation can be quickly completed with the reuse of the environments in the previous steps. 
Note that ${\tilde H}$ corresponds to the so-called {\it super-block} Hamiltonian in the DMRG. 
Diagonalizing ${\tilde H}$, we obtain the lowest eigenvalue and the corresponding eigenvector ${\tilde \Psi}_{abcd}^{~}$, which can be called the renormalized ground-state wavefunction. 

Figure\ \ref{fig:process_1} (b) shows the diagram for ${\tilde \Psi}_{abcd}^{~}$. 
Rearranging the leg indices into $ab$ and $cd$ in the matrix form, we next perform the singular-value decomposition (SVD) to obtain the new tensors
\begin{equation}
{\tilde \Psi}_{abcd}^{~} = \sum_{e}^{~} V_{ab}^{e} D_e^{~} V_{cd}^{e} \, ,
\label{Eq_2_5}
\end{equation}
where the r.h.s. is diagrammatically illustrated in Fig.\ \ref{fig:process_1} (c). 
Discarding tiny singular values, we keep the number of degrees of freedom for the new central bond $e$ within the upper bound $\chi$. 
If there is degeneracy in $D_{e}^{~}$, we keep all the degenerating singular values or discard all of them to maintain the symmetry of the state. 
Note that any basis truncation is not performed near the boundary of the TTN where the number of positive $D_{e}^{~}$ does not exceed $\chi$. 

After the decomposition in Eq.~(\ref{Eq_2_5}), the tensor $V_{be}^{a}$ in Fig.\ \ref{fig:process_1} (a) is replaced by $V_{ab}^{e}$ in Fig.\ \ref{fig:process_1}(c), where the position of the singular value is also shifted to the bond $e$. 
Also the isometry $V_{cd}^{e}$ can be updated.
Using the above local update accompanying the position shift of the central area, we sweep the entire TTN with improving the isometries. 
After repeating this procedure several times, we finally obtain a good variational wavefunction.

\section{Tree Structure Optimization}\label{sec:tree_str_opt}

\begin{figure}
\begin{center}
\includegraphics[width = 75 mm]{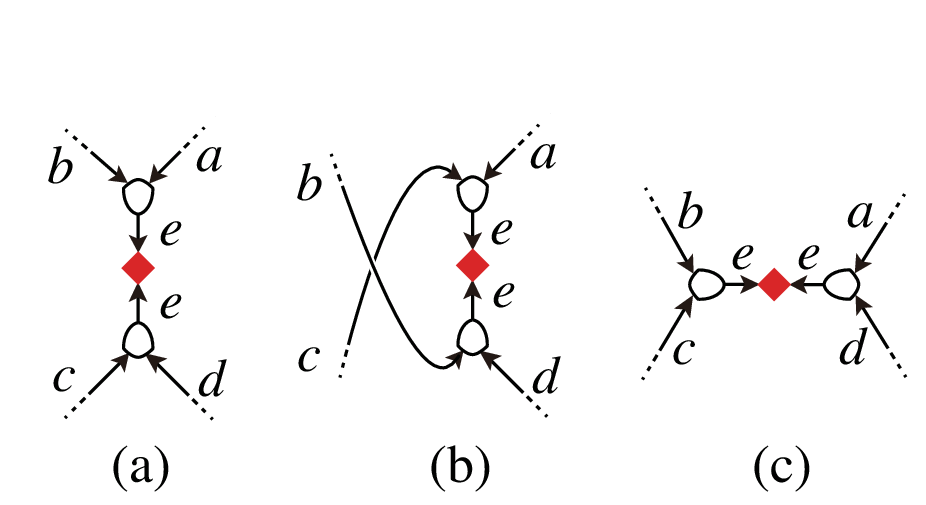}
\caption{Three possible applications of SVD to ${\tilde \Psi}_{abcd}^{~}$, as candidates for
the local reconnection.}
\label{fig:process_2}
\end{center}
\end{figure}

Now, we explain the variational scheme of automatically searching for the optimal TTN structure based on the conventional variational TTN algorithm above.
An essential idea is that the local reconnection of bonds can be implemented in every improvement step of isometries inside the central area. 
Let us see ${\tilde \Psi}_{abcd}^{~}$ in Fig.\ \ref{fig:process_1} (b) again. 
In Eq. (\ref{Eq_2_5}), we arranged the collective indices of $ab$ and $cd$ according to the original network structure and then performed SVD, as shown in Fig.\ \ref{fig:process_1} (c).
We draw this diagram again in Fig.\ \ref{fig:process_2} (a). 
As in Figs.\ \ref{fig:process_2} (b) and (c), however, there are the other two possible arrangements of the collective indices\cite{Vidal_2006}. 
If we adopt $ac$ and $bd$, the SVD yields
\begin{equation}
{\tilde \Psi}_{abcd}^{~} = \sum_{e}^{~} V_{ac}^{e} {D'}_{\!\!\! e}^{~} \! V_{bd}^{e} \, ,
\label{Eq_3_1}
\end{equation}
as shown in Fig.\ \ref{fig:process_2} (b). 
If we choose $ad$ and $bc$, we obtain
\begin{equation}
{\tilde \Psi}_{abcd}^{~} = \sum_{e}^{~} V_{ad}^{e} {D''}_{\!\!\!\! e}^{~} \! V_{bc}^{e} \, ,
\label{eq:caseC}
\end{equation}
which is depicted as Fig.\ \ref{fig:process_2} (c). 
These diagrams provide three options of the local connectivity of isometries in the TTN. 
In particular, Figs.\ \ref{fig:process_2} (b) and (c) accompany the reconnection of the network. 
Here, note that $D_e^{~}$ in Eq.~(\ref{Eq_2_5}), ${D'}_{\!\!\! e}^{~} \!\!$ in Eq.~(\ref{Eq_3_1}), and ${D''}_{\!\!\!\! e}^{~} \!\!$ in Eq.~(\ref{eq:caseC}) are different, but for all cases, the bond dimension of $e$ is at most $\chi$ after the basis truncation. 

In order to quantitatively evaluate ``quality" of the above three connections, we employ the EEs associated with their singular values.
Let us write the arrangement in Eq.~(\ref{Eq_2_5}) as $(ab|cd)$.
Then the corresponding EE is given by
\begin{eqnarray}
\mathcal{S}^{(ab|cd)}_{~} = - \sum_{e}^{~} \left( D_e^{~} \right)^2_{~} \ln \left( D_e^{~} \right)^2_{~} \, ,
\label{eq:EE}
\end{eqnarray}
where we have assumed the normalization of the TTN. 
Similarly, we have the EEs
\begin{eqnarray}
&& \mathcal{S}^{(ac|bd)}_{~} = 
- \sum_{e}^{~} \left( {D'}_{\!\!\! e}^{~} \!\! \right)^2_{~} \ln \left( {D'}_{\!\!\! e}^{~} \!\! \right)^2_{~} \, ,
\\
&& \mathcal{S}^{(ad|bc)}_{~} = 
- \sum_{e}^{~} \left( {D''}_{\!\!\!\! e}^{~} \!\! \right)^2_{~} \ln \left( {D''}_{\!\!\!\! e}^{~} \!\! \right)^2_{~} \, ,
\end{eqnarray}
for $(ac|bd)$ and $(ad|bc)$, respectively.
We then select the connection having the least EE as locally optimal connectivity of the isometries. 
Recall that in the TTN, cutting a certain bond always has the corresponding bipartitioning of the entire wavefunction. 
The choice of the reconnection, which locally satisfies the least-EE principle discussed in Sec.\ \ref{sec:Intro}, is thus expected to reduce the loss of accuracy caused by the limitation of finite bond dimension.
This reconnection process is the heart of the proposed algorithm.

\begin{figure}
\begin{center}
\includegraphics[width = 35 mm]{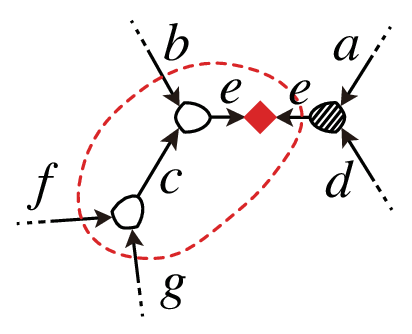}
\caption{The new central area in the case that Fig.\ \ref{fig:process_2} (c) is chosen as the optimal 
connection, and that $c$ is treated as the central bond of the new step. The shaded oval represents 
$V_{ad}^{e}$, which has been updated.}
\label{fig:process_3}
\end{center}
\end{figure}

Once the new connectivity inside the central area is determined, we also update the corresponding isometry.
Suppose that the connection of $(ad|bc)$ is selected, and the central bond is shifted to $c$ in the next step, as in Fig\ \ref{fig:process_3}.
Following Eq.~(\ref{eq:caseC}), we first update the isometry $V_{ad}^{e}$,  which is on the outside of the new central area illustrated as the dotted curve in Fig\ \ref{fig:process_3}. 
We next recalculate the block Hamiltonian and the spin operators belonging to $V_{ad}^{e}$, which turns out to be a part of ${\tilde H}$ in the next step. 
Meanwhile, we may skip the update of $V_{bc}^{e}$ since it is included in the new central area that will be treated in the next step.
Now, we can perform the same updating processes by shifting the central area.
The iterative algorithm summarized in Table\ \ref{Table_I} enables us to simultaneously optimize the variational energy $E_{\rm var}^{~}$ and the TTN structure by sweeping the central area over the entire TTN.

Here, it is worthwhile mentioning an acceleration algorithm effective for the numerical diagonalization of ${\tilde H}$.
So far, we have stored the wavefunction ${\tilde \Psi}_{abcd}^{~}$ and the updated isometry $V_{ad}^{e}$ in computer memory.
Also, we assume that the isometry $V_{fg}^{c}$ was obtained previously. 
Then, we calculate the contraction $\sum_{acd}^{~} V_{fg}^{c} {\bar V}_{ad}^{e} \, {\tilde \Psi}_{abcd}^{~}$, which can be a good initial vector for ${\tilde \Psi}_{bfge}^{~}$.
Alternatively, if we have $V_{bc}^{e}$ and ${D''}_{\!\!\!\! e}^{~}$ explicitly, the initial vector can be constructed as
$\sum_c^{~} V_{fg}^{c} V_{bc}^{e} \, {D''}_{\!\!\!\! e}^{~}$.
These initial vectors may substantially reduce the computational cost of the Lanczos or the Davidson method to diagonalize ${\tilde H}$ in the next step \cite{White_1996}.

\begin{table}[b]
\caption{Algorithm of automatic structural optimization for the TTN.}
\label{Table_I}
\begin{center}
\begin{tabular}{ll}
\hline
\hline
1. & Construct the super-block  Hamiltonian ${\tilde H}$ according to\\
   & the current central area. \\
2. & Diagonalize ${\tilde H}$ to obtain the renormalized ground-state\\
   & wavefunction ${\tilde \Psi}$. \\
3. & Perform the SVD on ${\tilde \Psi}$ in three different manners as\\
   & shown in Fig.\ \ref{fig:process_2}. \\
4. & Choose the grouping with the smallest EE and update \\
   & the tensors in the central area accordingly. \\
5. & Move the central bond to the appropriate direction \\
   & (see Appendix \ref{App:Sweep}). \\
6. & Iterate the above steps until the entire TTN is updated. \\
\hline
\hline
\end{tabular}
\end{center}
\end{table}

In the following, we describe several comments on the technical aspect of the algorithm in order.
The initial setup of the TTN is flexible since both the tree structure and the isometries will be optimized afterward. 
A MPN generated by the finite-system DMRG is one of the realistic candidates. 
It is even possible, in principle, to use a naive product state, which can be regarded as a TTN with $\chi = 1$. 
However, a good initial setup may decrease the number of sweeps required for convergence. 
For random spin systems, the TTN generated by the tSDRG protocol\cite{HikiharaFS1999,GoldsboroughR2014,LinKao_2017,SekiHO2020,SekiHO2021,Ferrari_2022} will be a reasonable initial TTN. 
Also, we note that the central area must move around all auxiliary bonds and isometries in the sweeping process of the network. 
In Appendix \ref{App:Sweep}, we explain the computational detail of this technical issue about the sweeping path of iterative computation.

While we employ the bond EE for the evaluation of the local network structure in the proposed algorithm, one may use other quantities such as the truncation error, i.e., the sum of the eigenvalues of the reduced DM for the discarded states, or the Renyi entropy\cite{LiRYS2022}.
%A candidate is the truncation error, that is, the sum of the eigenvalues of the reduced DM for the discarded states.
The truncation error is directly related to the variational energy, while the bond EE can access the entanglement structure of the target state.

Since the reconnection of the isometries in the central area is always local, the optimization process of the TTN structure may be trapped at local minima of the distribution landscape of the EEs on the network. 
A possible device to escape from such trapping is to combine a stochastic method with the algorithm.
More precisely, we can select one of the diagrams in Fig.\ \ref{fig:process_2} according to the relative probabilities based on the heat-bath method,
\begin{subequations}
\label{Eq:Prob_SVD_options}
\begin{align}
P^{(ab|cd)}_{~} \propto \exp\left[ - \beta \mathcal{S}^{(ab|cd)}_{~} \right] \, , 
\label{Eq:Prob_SVD_abcd} \\
P^{(ac|bd)}_{~} \propto \exp\left[ - \beta \mathcal{S}^{(ac|bd)}_{~} \right] \, , 
\label{Eq:Prob_SVD_acbd} \\
P^{(ad|bc)}_{~} \propto \exp\left[ - \beta \mathcal{S}^{(ad|bc)}_{~} \right] \, ,
\label{Eq:Prob_SVD_adbc}
\end{align}
\end{subequations}
where $\beta$ is an effective inverse temperature. 
If $\beta$ is gradually increased during the sweeping process, the TTN structure converges to the one where the sum of EE on all the bonds is minimized, which is a candidate for the optimal TTN.
%During the sweeping process, $\beta$ is gradually increased, and finally, the TTN structure is converged to the optimal one. 
%In principle, this annealing process also minimize the sum of EEs on all the bonds in the TTN. 
If one prefers to suppress the EE on each bond further, the square or higher power of $\mathcal{S}^{(ab|cd)}_{~}$, $\mathcal{S}^{(ac|bd)}_{~}$, and $\mathcal{S}^{(ad|bc)}_{~}$ can be used in the exponent of the r.h.s. of Eq.~(\ref{Eq:Prob_SVD_options}). 
Note that the stochastic sampling of the tree structure can be parallelized.

\section{Numerical results}\label{sec:numerics}

\begin{figure}
\begin{center}
\includegraphics[width = 60 mm]{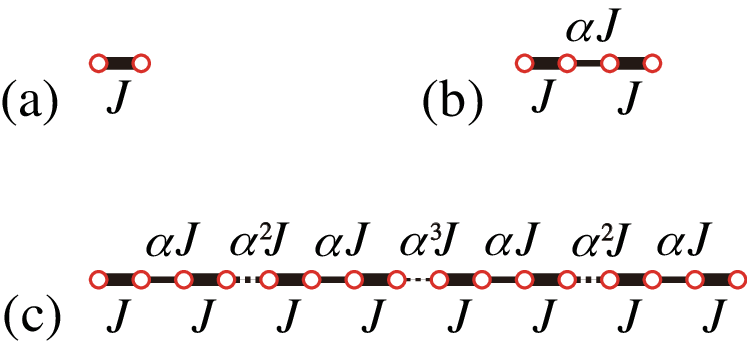}
\caption{Inhomogeneous interactions on the hierarchical chain when (a) $N = 2$, (b) $N = 4$, 
and (c) $N = 16$.}
\label{fig:Hier_chain}
\end{center}
\end{figure}

We check the validity of the proposed optimization scheme in Table\ \ref{Table_I} by applying it to the inhomogeneous Heisenberg spin chain, whose Hamiltonian is formally written as
\begin{eqnarray}
{\hat H} = \sum_{i=1}^{N-1} J_i^{~} \, {\bm S}_i^{~} \cdot {\bm S}_{i+1}^{~},
\label{eq:Hier_chain}
\end{eqnarray}
where ${\bm S}_i^{~}=(S^x_i, S^y_i, S^z_i)$ represents the $S = 1/2$ quantum spin on the $i$-th site.
We consider the case where the position-dependent exchange coupling $J_i^{~} > 0$ is recursively determined as shown in Fig.\ \ref{fig:Hier_chain}.
We call the model the hierarchical chain in the following. 
The minimum unit is a two-site system ($N = 2$) in Fig.\ \ref{fig:Hier_chain} (a), where the coupling constant between ${\bm S}_1^{~}$ and ${\bm S}_2^{~}$ is $J > 0$, which defines the unit of energy scale.
Joining two two-site units with the coupling $\alpha J$, we obtain a four-site unit ($N = 4$) in Fig.\ \ref{fig:Hier_chain} (b).
The parameter $0 < \alpha \le 1$ controls the decay rate of the coupling. 
In general, joining the $2^n_{~}$-site units ($N = 2^n_{~}$) with the coupling $\alpha^{n}_{~} J$, we obtain the $2^{n+1}_{~}$-site system. 
Figure\ \ref{fig:Hier_chain} (c) shows the $2^4_{~}$-site system as an example.
We treat the system size $N = 64$ (for several $\alpha$'s) and $N = 128$ (for $\alpha=0.50$) in the following numerical calculations.

A significant feature of the hierarchical chain is that when $\alpha$ is sufficiently small, one can deduce its optimal TTN with the perturbative RG scheme.
In the hierarchical chain, the largest exchange coupling is $J$, with which two spins ${\bm S}_{2l-1}$ and ${\bm S}_{2l}$ are entangled most strongly.
Therefore, it is natural to connect the two spins coupled with $J$ by an isometry to form the spin blocks at the first step of RG.
In the second step, two blocks connected via $\alpha J$ are entangled most strongly, and thus one should put an isometry to merge the blocks to form a new block.
A recursive application of the above RG process eventually gives rise to the perfect binary TTN, which is expected to be the optimal one as long as $\alpha$ is small enough.
As $\alpha$ increases to unity, the system approaches the uniform Heisenberg chain, where the critical ground state is realized. 
Thus, the hierarchical chain may provide a good platform to visualize the crossover of the entanglement structure between the ones of the perfect binary tree and of the uniform wavefunction with critical fluctuation.

In order to demonstrate that the optimal TTN can be automatically obtained with the proposed algorithm, we start the calculation from the MPN prepared by the following processes. 
We first focus on two spins at an open end of the chain.
We then diagonalize the block Hamiltonian, truncate the high-energy eigenstates, and include the neighboring spin into the new block. 
We perform this recursive RG process from both ends of the chain and increase the block size one by one.
When the left and right blocks meet at the center, we have an MPN including all the isometries, from which we start the first sweep of the structural optimization algorithm.
The upper bound of the bond dimension in the initial MPN is $\chi = 40$.

During the iterative sweeps of the algorithm in Table\ \ref{Table_I}, the isometries and their connectivity is successively modified.
We perform the diagonalization of the effective Hamiltonian ${\tilde H}$ 
within the subspace of zero total 
magnetization $\sum_i S^z_i = 0$ and take the lowest-energy state of the subspace as the ground state.
After the numerical convergence, the variational state with optimal TTN structure is 
automatically obtained. For the hierarchical chain, we do not use the stochastic choice of reconnection 
in Eq.~(\ref{Eq:Prob_SVD_options}), but always choose the smallest EE connection in each local update. We perform the SVD of the 
renormalized ground-state wavefunction ${\tilde \Psi}$ by the diagonalization 
of the reduced density matrix (DM), whose eigenvalues are equal to the square of the corresponding 
singular values. 
We note that the SVD can also be directly carried out by using a linear algebra package.
For comparison, we performed additional calculation with forbidding the reconstruction of TTN structure, which remains MPN throughout the variational calculation. Note that this calculation is equivalent to 
the finite-system DMRG method.

We set the maximum number of the bond dimension to be $\chi = 40$. We discard the 
block states whose DM eigenvalues are smaller than $10^{-12}$, since their contribution to the 
ground-state wavefunction is negligible. Due to this cut-off condition, the number of block states 
kept in the practical calculation can be smaller than $\chi$. 
For example, we kept only $26$ states, at most, for $\alpha = 0.5$ 
and $N = 64, 128$, while $40$ states were kept for $\alpha = 1.0$ and $N = 64$. The truncation error 
%, the sum of the DM weights of the discarded states, 
is less than $1.3 \times 10^{-11}$ for $(\alpha,N)=(0.50,64), (0.50,128), (0.75,64), (0.80, 64)$ and $2.5 \times 10^{-8}$ at the worst case of $(\alpha, N)=(1.00, 64)$.

\begin{figure}
\begin{center}
\includegraphics[width = 75 mm]{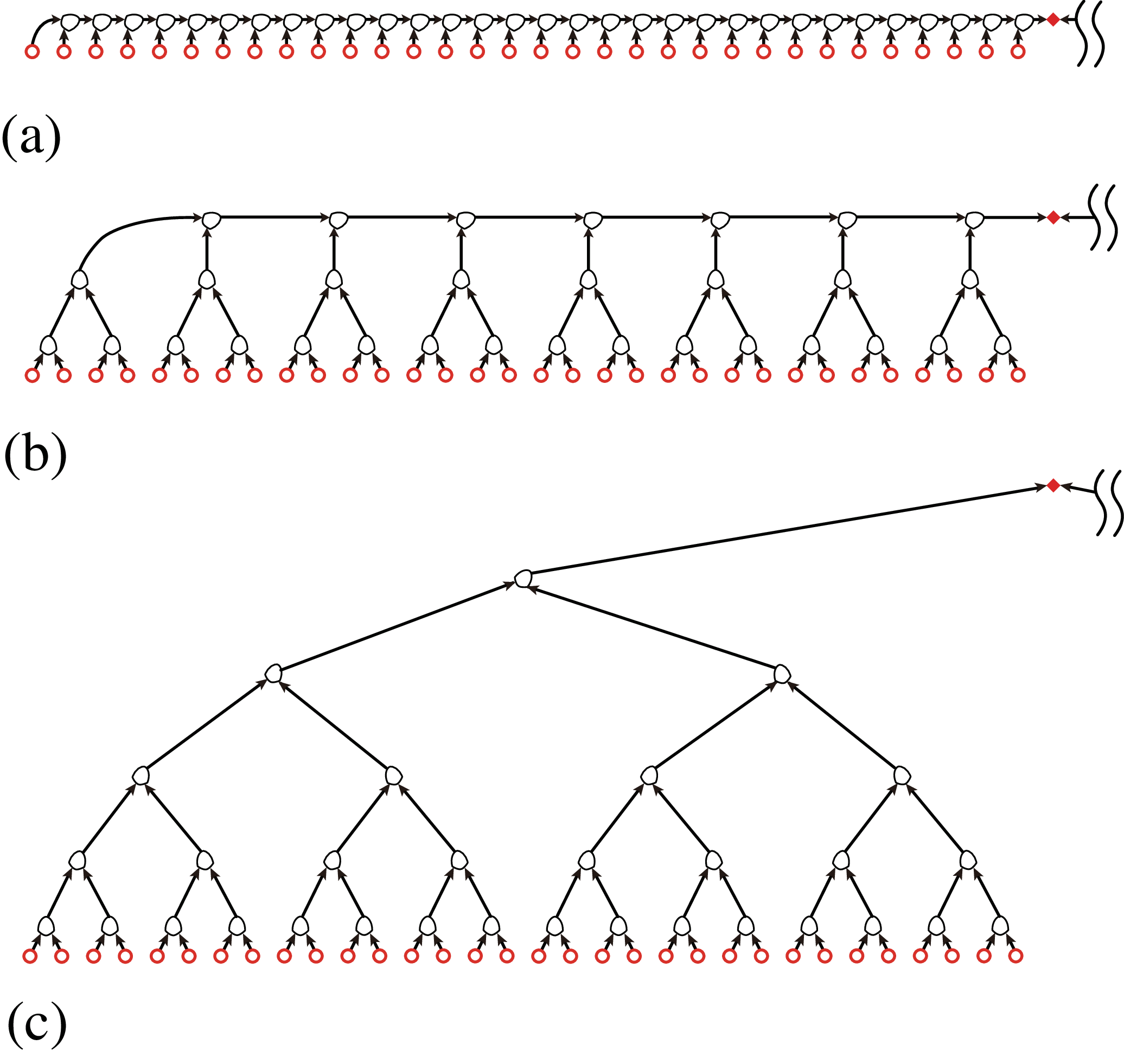}
\caption{Optimization process when $\alpha = 0.5$ and $N = 64$. 
(a) The initial MPN.  (b) After the first sweep. (c) Any time after the second sweep. 
Only the left half of the TTN is presented while the right half is symmetric 
with respect to the center of the system.
}
\label{fig:Hier-trans}
\end{center}
\end{figure}

Figure\ \ref{fig:Hier-trans} shows the transition process of the TTN structure during the sweeps for $\alpha = 0.5$ and $N = 64$. 
The calculation starts with the initial 
MPN shown in Fig.\ \ref{fig:Hier-trans} (a). After the first sweep, perfect-binary-tree-like structure 
appears for every four sites, as shown in Fig.\ \ref{fig:Hier-trans} (b), where the MPN structure remains 
in the upper part of the diagram. After the second sweep, the perfect binary tree structure shown in 
Fig.\ \ref{fig:Hier-trans} (c) expands up to the top. This is the optimal TTN structure, which is unchanged afterward.
We emphasize that the optimal TTN structure obtained by the calculation agrees with the one derived from the perturbative RG scheme discussed above.
We observed the similar transition process for $\alpha = 0.5$ and $N = 128$, 
where the perfect binary TTN emerges after three sweeps. 
Here, we note that the intermediate TTN structure may depend on how the center bond moves during a sweep;
the structure in Fig.\ \ref{fig:Hier-trans} (b) is the one obtained with the sweep procedure explained in Appendix\ \ref{App:Sweep}.
The point is that as the calculation proceeds, the TTN structure smoothly changes toward the optimal one shown in Fig.\ \ref{fig:Hier-trans} (c).

\begin{table}[b]
\caption{Maximum and average of the bond EE obtained for optimized TTN and fixed MPN when 
$N = 64$ and $128$ with $\alpha = 0.5$. The EEs on the boundary bonds are not included in the analysis.}
\label{Table_II}
\begin{center}
\begin{tabular}{l r r}
\hline
\hline
Type &  ~~~Maximun & ~~~~Average \\
\hline
optimized TTN ($N = 64$) & 0.1110 & 0.0640 \\
optimized TTN ($N = 128$) & 0.1110 & 0.0625 \\
MPN ~ ($N = 64$) & 0.6935 & 0.3697 \\
MPN ~ ($N = 128$) & 0.6935 & 0.3719 \\
\hline
\hline
\end{tabular}
\end{center}
\end{table}

Table\ \ref{Table_II} shows the maximum and the average of the bond EE obtained in the optimized TTN for $\alpha = 0.5$ and $N = 64, 128$. For comparison, 
the same set of data obtained in the fixed MPN are shown. Here, the bonds directly connected to the bare spins are not included in the analysis, 
since they always carry the EE of the amount $\ln 2$ irrespective of the network structure, reflecting the fact that the two states 
$|\uparrow \, \rangle$ and $|\downarrow \, \rangle$ of each bare spin have the same reduced-DM 
weight $1/2$. It is clearly seen in the table that both the maximum and average of the bond EEs are lessened by the tree structure optimization, with the suppression ratio about $1/6$.

\begin{figure}
\begin{center}
\includegraphics[width = 80 mm]{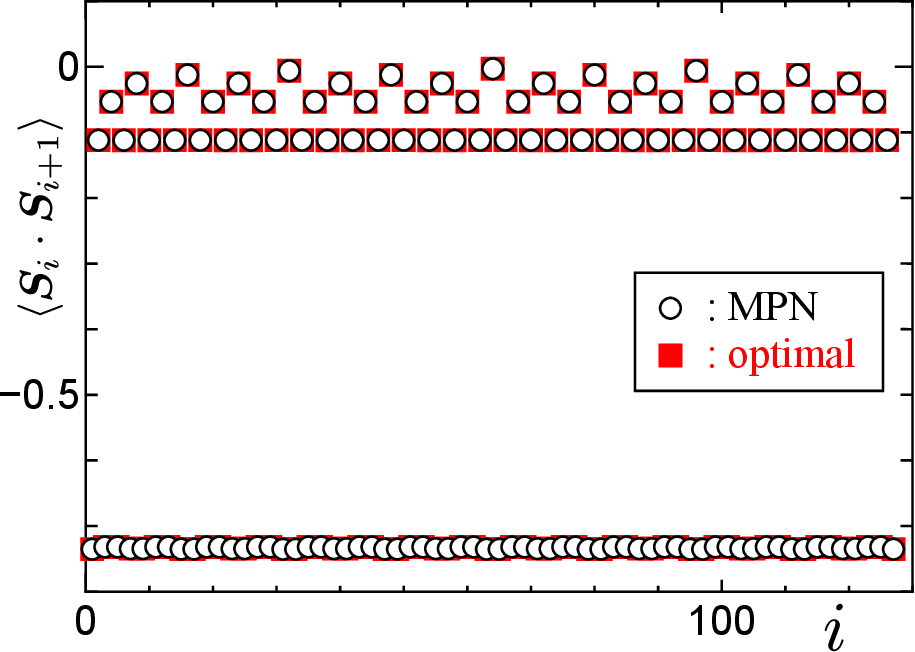}
\caption{Nearest-neighbor spin correlation function when $\alpha = 0.5$ and $N = 128$. 
Open circles and solid squares, respectively, represent the obtained result by the MPN and 
the optimal TTN. 
}
\label{fig:Hier-NNcor}
\end{center}
\end{figure}

Figure\ \ref{fig:Hier-NNcor} shows the expectation values of the nearest-neighbor spin-correlation functions $\langle {\bm S}_{i}^{~} \cdot {\bm S}_{i+1}^{~} \rangle$ when $\alpha = 0.5$ and $N = 128$.
As seen in the figure, the correlation functions obtained from the MPN coincide with those from the optimal TTN. 
This is because the ground state for $\alpha=0.5$ is close to the product state consisting of the singlet dimers, and accordingly, both the optimized TTN and MPN obtained have the sufficient potential of representing the ground state accurately enough.
Indeed, the strong dimerization in every two sites can be confirmed in the figure where $\langle {\bm S}_{2\ell-1}^{~} \cdot {\bm S}_{2\ell}^{~} \rangle$ is distributed within the range $-0.732 \sim -0.735$, while the spin correlations across neighboring dimers are weak. 
Also, we note that the ground-state energy calculated by the MPN agrees with that by the optimal TTN up to $11$ digits.

\begin{figure}
\begin{center}
\includegraphics[width = 80 mm]{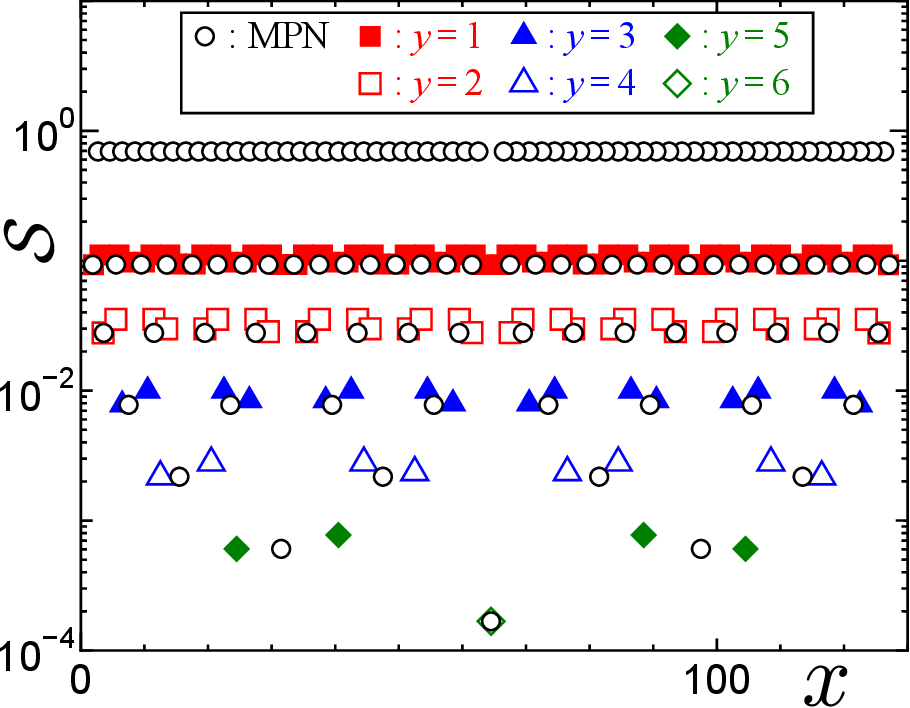}
\caption{Bond EE calculated when $\alpha = 0.5$ and $N = 128$ as a function of the horizontal position of the bonds, $x$, defined in the text. 
Open circles represent the EE calculated by the MPN. 
Other plots are calculated by the optimal TTN. 
$y$ denotes the height of the bond in the optimal TTN.
}
\label{fig:Hier-EE}
\end{center}
\end{figure}

Distinctions between the optimized TTN and the MPN can be found in the spatial distribution of the EE on the auxiliary bonds, although both approaches provide the accurate results at the level of the correlation functions. 
In Fig.\ \ref{fig:Hier-EE}, we compare the EE distributions on the two different networks of the optimized TTN and the MPN for $\alpha = 0.5$ and $N = 128$.
In the figure, we have introduced the horizontal coordinate $x$ (for both the optimized TTN and the MPN) and the hight coordinate $y$ (only for the optimized TTN) to specify the positions of the auxiliary bonds;
The definitions of the coordinates are presented in Appendix\ \ref{App:coordinate}.
%We first assign the site index $i$ to the horizontal coordinate of bare spins.
%The position of an isometry is then given by the average of the horizontal coordinates of bare spins or the isometries connected with the incoming bonds.
%The position $x$ of the auxiliary bonds is finally defined as the average of the coordinates of the connected isometries. 
%For the case of the optimized TTN, moreover, we have introduced the height coordinate $y$ for the position of bonds, which is defined as the distance of the bond from the nearest bare spin.
The EEs on the boundary bonds directly connected to the bare spins, that take the value $\mathcal{S} = \ln 2$, are not included in Fig.\ \ref{fig:Hier-EE}. 

Let us discuss the distribution of the EE in Fig.\ \ref{fig:Hier-EE} from the bottom to the top.
First, the same smallest value of $\mathcal{S}=1.7\times 10^{-4}$ at $x=64.5$ is found in the data of the optimized TTN and the MPN, since the corresponding bipartition of the system into the left half $1 \le i \le 64$ and the right one $65 \le i \le 128$ is captured by both networks.
Also, both of them capture the second smallest $\mathcal{S}=6.1 \times 10^{-4}$ at $x=24.5, 104.5$ for the optimized TTN and $x \simeq 31.5, 97.5$ for the MPN, corresponding to the bipartition into $1 \le i \le 32$ and $33 \le i \le 128$ and into $1 \le i \le 96$ and $97 \le i \le 128$. 
Meanwhile, only the optimized TTN captures a slightly larger value of $\mathcal{S}=7.7 \times 10^{-4}$ at $x=40.5$ and $88.5$, which corresponds to the bipartitions into $33 \le i \le 64$ and the rest and into $65 \le i \le 96$ and the rest.
Note that these bipartitions separate an inner branch of the perfect-binary-tree network from the rest.
The MPN does not have the bond corresponding to such bipartitions and thus can not capture the EE of $\mathcal{S}=7.7 \times 10^{-4}$.
As the hight coordinate $y$ in the optimized TTN decreases, the number of the bipartitions with small EEs that cannot be captured by the MPN increases.
In the upper part of the figure, the EEs on almost half of the auxiliary bonds in the MPN are take values slightly larger than $\ln 2$, since the bonds carry the EE of the strongly-dimerized pairs sitting in every two sites.
Such large EEs never appear in the optimized TTN, indicating that the automatic structural optimization algorithm successfully avoids the bipartitions with the large EEs.

\begin{figure}
\begin{center}
\includegraphics[width = 75 mm]{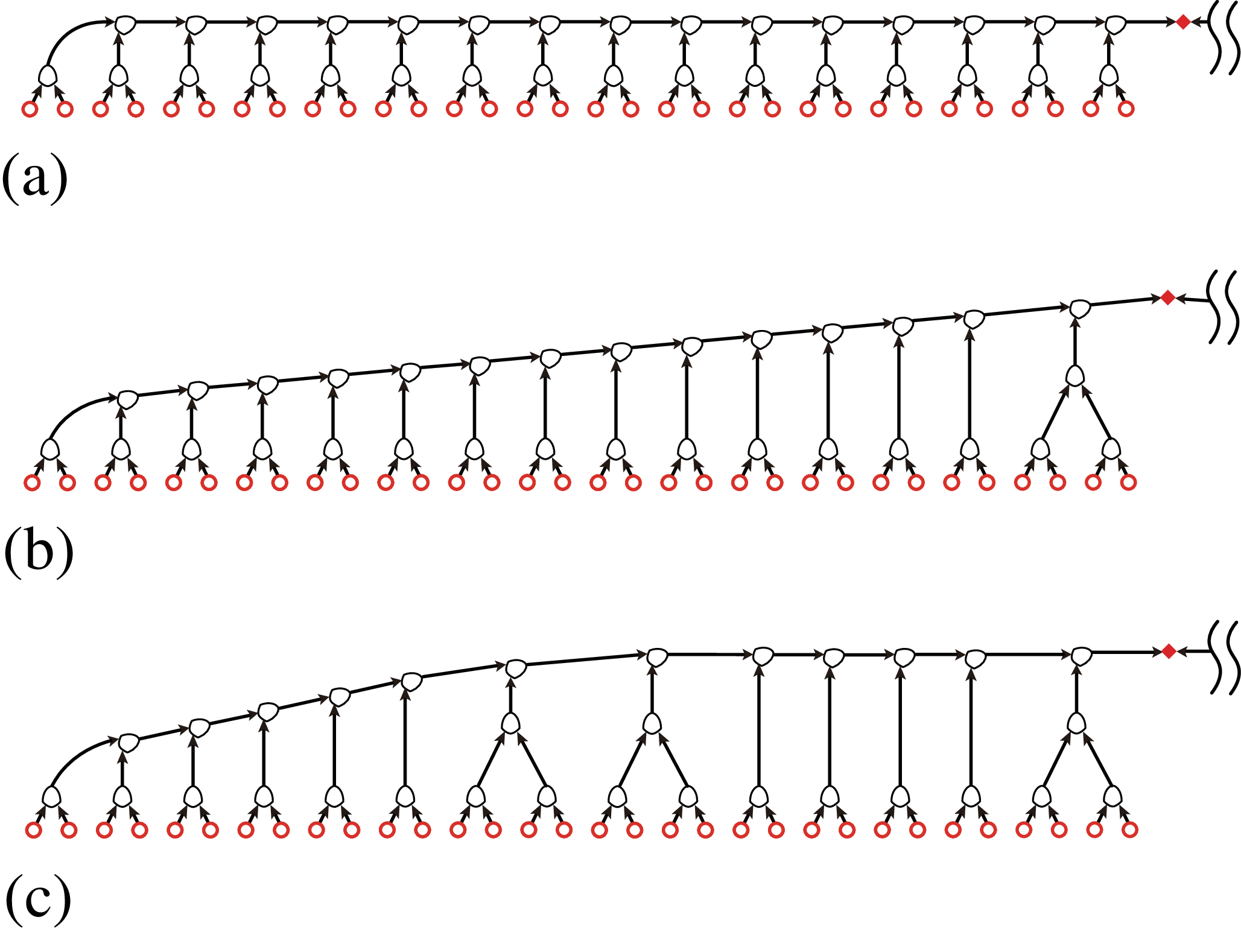}
\caption{Optimal structures for (a) $\alpha = 1.00$, (b) $\alpha = 0.80$, and (c) $\alpha = 0.75$. 
The system size is $N = 64$. 
Only the left half of the TTN is presented while the right half is symmetric 
with respect to the center of the system.
}
\label{fig:trans}
\end{center}
\end{figure}

Finally, we discuss the $\alpha$ dependence of the optimal tree structures. 
Figure\ \ref{fig:trans} (a) shows the optimized TTN for the chain with $\alpha = 1$ and $N = 64$, which is nothing but a uniform Heisenberg chain. 
In the bulk limit, the ground state of the chain is usually described by the uniform MPS.
In the finite-size system with open boundaries, however, the strong dimer instability induces a sizable staggered component both in the nearest-neighbor spin correlations and the EEs\cite{LaflorencieSCA2006}.
Reflecting this dimerization effect, the optimized TTN for $\alpha=1$ and $N = 64$ has the MPN structure in the unit of the isometries merging every pair of neighboring bare spins. 
As $\alpha$ decreases, the hierarchical distribution of the EE gradually enhances between the dimerized MPN for $\alpha = 1$ and the perfect binary tree for $\alpha = 0.5$ [Fig.\ \ref{fig:Hier-trans} (c)].
In Fig.\ \ref{fig:trans} (b) and (c), we respectively show the optimal TTN structures for $\alpha = 0.8$ and $0.75$ with $N = 64$. 
Here, recall that in the hierarchical chain, the weakest coupling is placed at the center of the chain and the second weakest at the quarter from the open boundary.
Figures\ \ref{fig:trans} (b) and (c) clearly illustrate that the perfect-binary-tree-like structure grows from the regions around the weaker interactions.

\section{Concluding remarks}\label{sec:conc}

We have proposed the automatic structural optimization algorithm for the TTN, where the auxiliary bonds between isometry tensors are locally reconnected based on the least-EE principle during the variational sweeping on the TTN.
We have then demonstrated that for the hierarchical chain, the algorithm generates a perfect binary tree as the optimal network structure, where the emergence of the bonds with a large EE was actually suppressed. 
The resulting network structure can be regarded as a fingerprint of the entanglement structure contained in the quantum state. 

In this paper, we have focused on the analysis of the distribution of the EE in the optimal TTN to verify that the proposed algorithm indeed works well in terms of the least-EE principle.
In practical situations, of course, it is also an important issue to clarify how the variational energy $E_{\rm var}^{~}$ in Eq.~(\ref{Eq_2_4}) can be improved with the structural optimization. 
Preliminary calculations for random quantum spin systems with long-range interactions suggest that the optimal TTN realized in the proposed algorithm provides a better accuracy of $E_{\rm var}^{~}$ compared with the MPN in the DMRG and a TTN obtained by the tSDRG. 
For such complex systems, the annealing process according to Eq.\ (\ref{Eq:Prob_SVD_options}) is essential to escape from the local minimum in the EE landscape.
The detailed analysis for the random spin system will be published elsewhere.

The structural optimization based on the local reconnection scheme is promising to be effective in a wide variety of TTN applications to, {\it e.g.}, two- or higher-dimensional systems and systems with complicated interactions.
In particular, random spin systems, for which the optimal TTN is expected to exhibit non-trivial structure, are interesting subject for the application of the present algorithm.
Quantum chemistry systems containing complex configuration interactions among electrons\cite{Nakatani_2013,Murg_2010,Murg_2014,Gunst_2018,Gunst_2019,Larsson2019,LiRYS2022,ChanHG2002,MoritzHR2005} are another intriguing candidate to be studied.
How the structural optimization of the TTN improves the computational accuracy of such systems is an interesting future problem.
In addition, the real-time evolution of quantum many-body states is often described with the framework of the TTN\cite{Vidal_2006,Bauernfeind_2020,Cao_2021}.
Emulation of quantum circuits can be also viewed as a sort of the time-evolved quantum state generated by gate operations and measurements\cite{Seitz_2022}. 
The dynamical reconnection of the TTN for such time evolution problems is also a potential application target.

For the practical application of the algorithm, the estimation of computational costs is another imporant issue.
In the variational TTN calculation, the most time consuming part is the diagonalization of the renormalized Hamiltonian ${\tilde H}$ to obtain the renormalized ground-state wavefunction ${\tilde \Psi}$. 
The cost scales as $O(N_{\rm int} \chi^5)$, where $N_{\rm int}$ is a numerical factor that depends on the boundary area or volume of subsystems connected to the central area and the range of interaction.
%of the order of the number of interactions between the four subsystems.
(See Appendix\ \ref{App:Comp_cost} for the evaluation of the cost.)
Another computational hotspot with the cost $O(\chi^6)$ is the full diagonalization of the reduced DM.
The cost for this part may be reduced to $O(\chi^5)$ or $O(\chi^4 \log\chi)$ by use of the partial SVD or randomized SVD~\cite{Halko_2011}.
These computational costs in the proposed algorithm based on the TTN are much heavier than those in the DMRG based on the MPN since in the DMRG two of the four bonds entering the central area always have the dimension $\nu$, the degree of freedom of a bare spin, and the relation $\chi \gg \nu$ usually holds.

A possible scheme to reduce the computational cost is to carry out the structural optimization with a small $\chi$, and then improve isometries in the obtained optimal TTN structure with a large $\chi$.
This two-step optimization is effective if the convergence of the TTN structure with respect to $\chi$ is fast. 
Variational improvement of the isometries can also be accelerated by employing the single-site update algorithm\cite{Takasaki_1999,White_2005}, in which the central area contains only a single isometry.
The dimension of ${\tilde H}$ is thus suppressed to be of the order of $\chi^3$ so that the cost of diagonalizing ${\tilde H}$ becomes $O(N_{\rm int} \chi^4)$.
If the system possesses a high symmetry, the use of the symmetric bases~\cite{McCulloch_2002} would be further efficient.
The techniques in the real-space parallel DMRG~\cite{Stoudenmire_2013}, that are effective in accelerating the sweep process for large systems, can also be incorporated with the present algorithm.

Finally, we would like to comment on structural optimization for tensor networks other than the TTN.
In general, the philosophy of optimal network structure with the least-EE principle can be relevant to a wider class of tensor networks containing loop structure.
In particular, how to clarify the optimal MERA structure may be an interesting problem in connection with the current study, since the MERA network is constructed by inserting disentanglers into the TTN.

\begin{acknowledgments}
This work is partially supported by
% JSPS 
KAKENHI Grant Numbers JP19K03664, JP20K03766, JP21K03403, JP22H01171, JP21H04446, 
and a Grant-in-Aid for Transformative Research Areas ``The Natural Laws of Extreme 
Universe---A New Paradigm for Spacetime and Matter from Quantum Information" 
(KAKENHI Grant Nos. JP21H05182, JP21H05191) from JSPS of Japan.
It is also supported by JST PRESTO No. JPMJPR1911, MEXT Q-LEAP Grant No. 
JPMXS0120319794, and JST COI-NEXT No. JPMJPF2014.
H.U and T.N was supported by the COE research grant in computational science from Hyogo 
Prefecture and Kobe City through Foundation for Computational Science.
\end{acknowledgments}

\appendix
\section{Sweep procedure}\label{App:Sweep}

The automatic optimization scheme we have explained performs local reconnection of isometries, 
when it is necessary, therefore a computational care must be taken so that all the isometries 
are updated once, at least, during a sweep. We introduce the {\it flags} for all the bonds and isometries, 
in order to indicate whether they have been updated (flag on) or not (flag off). At the beginning of each 
sweep, all the boundary bonds that are directly connected to bare spins are set to be flag on, while other bonds and 
all the isometries are flag off. We may choose an arbitrary bond, excluding the boundary ones, as the 
{\it origin}, and start the sweeping by selecting a bond which is flag off and neighboring to the origin as the central one 
for the first update step. 

After the optimal local connection in the central area is determined in each step, we select a bond from 
flag-off ones entering to the current central area as the central bond in the next step. If there are 
multiple choices, we select the one which is most distant from the origin. 
If there is still more than one choice, one may choose any of them.
For each isometry in the 
current central area, if both the entering bonds are flag on, we turn on the flag of the isometry and that 
of the outgoing bond, which is the current central bond. 
Exception applies to the step when the central bond comes to the origin;
after the update of the isometry and local connection, we keep the flags off for the origin bond and the isometries connected to it even if the condition for turning on the flags is satisfied.

By means of these flag operations, the central bond maneuvers around in the manner that it 
moves towards the boundary of TTN, if possible, along the bonds which are flag off.
If the central bond hits a dead end of flag-on bonds, it steps back towards the origin with turning on the flags of the bond and isometry of the dead-end branch.
We finish the sweep when all the bonds neighboring to the central bond are flag on;
this situation occurs in the step where the central bond is at the origin.
(We note that even when the central bond comes to the origin, we continue the sweep if there remains a flag-off bond neighboring to the origin.)
At the end of the sweep, we update the isometries and local network connection around the origin. We perform the next sweep in the same manner.

We note that the procedure discussed above is not the only way to realize the adequate sweep.
The sweeping path of the central bond should affect the convergence speed of the calculation.
To find the best way of sweeping in practice will be a future problem.

\section{Computational cost}\label{App:Comp_cost}

The computational cost for the diagonalization of the renormalized Hamiltonian ${\tilde H}$ to obtain the renormalized ground-state wavefunction ${\tilde \Psi}$ scales with the cost to multiply ${\tilde H}$ to a vector $\tilde{\psi}$ in the Hilbert space of the superblock, which is expanded by the four bonds entering the central area.
Here, let us consider the general Heisenberg model in which any pair of spins may have the exchange interaction.
The configuration of a conventional DMRG-like superblock Hamiltonian, described for example in Fig.~\ref{fig:process_1}, is given by
\begin{equation}
\tilde{H}=\tilde{H}^{(ab)}_{0} + \tilde{H}^{(cd)}_{0} + \sum_{i \in (ab)} \sum_{i' \in (cd)} J_{i, i'} {\bm S}_i^{~} \cdot {\bm S}_{i'}^{~},~
\label{eq:decomposition_1}
\end{equation}
where $\tilde{H}^{(ab)}_{0}$ and $\tilde{H}^{(cd)}_{0}$ are renormalized Hamiltonians for the subsystems $(ab)$ and $(cd)$, respectively, and $J_{i, i'}$ is the exchange constant between spins in $i$th and $i'$th sites.
If one employs this decomposition, a computational cost of $O(\tilde{N}_{\rm int}\chi^6)$ is required for the operation of $\tilde{H}\tilde{\psi}$, where $\tilde{N}_{\rm int}$ is the number of operations to multiply the exchange interaction terms in Eq.\ (\ref{eq:decomposition_1}) to $\tilde{\psi}$.
For the systems with short-range interactions, only $J_{i,i'}$ for the spins close to the boundary between the subsystems $(ab)$ and $(cd)$ is nonzero so that $\tilde{N}_{\rm int}$ scales with the boundary area between the subsystems.
For the systems with long-range interactions, all $J_{i.i'}$ are nonzero in general.
In this case, one can rewrite the exchange interaction part in Eq.\ (\ref{eq:decomposition_1}) as $\sum_{i \in (ab)} {\bm S}_i^{~} \cdot \left(\sum_{i' \in (cd)} J_{i, i'} {\bm S}_{i'}^{~}\right)$ or $\sum_{i' \in (cd)} {\bm S}_{i'}^{~} \cdot \left(\sum_{i \in (ab)} J_{i, i'} {\bm S}_{i}^{~}\right)$ so that $\tilde{N}_{\rm int}$ scales with the volume of the smaller subsystem.
%\begin{equation}
%\tilde{H}=\tilde{H}^{(ab)}_{0} + \tilde{H}^{(cd)}_{0} + \sum_{\ell=1}^{\tilde{N}_{\rm int}} \tilde{O}^{(ab)}_{\ell} \tilde{O}^{(cd)}_{\ell},~
%\label{eq:decomposition_1}
%\end{equation}
%where $\tilde{H}^{(ab)}_{0}$ and $\tilde{O}^{(ab)}_{\ell}$ are a renormalized Hamiltonian and set of operators in the subsystems specified by bonds $a$ and $b$, respectively [the same for $\tilde{H}^{(cd)}_{0}$ and $\tilde{O}^{(cd)}_{\ell}$]. 
%Also $\tilde{N}_{\rm int}$ has about as many degrees of freedom as the number of interactions between the two subsystems and scales with the surface area/volume of the smaller of the two subsystems in volume when the system has short/long-range interactions.
%When we employ this decomposition, a computational cost of the order of $(2+2\tilde{N}_{\rm int})\chi^6$ is required for the operation of $\tilde{H}{\tilde \Psi}$.
%
Furthermore, one can reduce the order of computational cost with respect to $\chi$ by dividing the superblock into four subsystems: 
\begin{equation}
\tilde{H} = \sum_{p} \tilde{H}^{(p)}_{0} + \sum_{p} \sum_{p'(>p)} \sum_{i \in p} \sum_{i' \in p'} J_{i, i'} {\bm S}_i^{~} \cdot {\bm S}_{i'}^{~},
\label{eq:decomposition_2}
\end{equation}
where $p,p' \in \{ a,b,c,d\}$ and $\tilde{H}^{(p)}_{0}$ is a renormalized Hamiltonian for the subsystem $p$.
In this decomposition, the cost to multiply $\tilde{H}$ to $\tilde{\psi}$ is $O(N^{~}_{\rm int} \chi^5)$, where $N^{~}_{\rm int}$ is the number of operations to multiply the exchange interaction part in Eq.\ (\ref{eq:decomposition_2}).
$N^{~}_{\rm int}$ scales with the boundary area between (the volume of the smaller one of) two out of the four subsystems in the case of short-range (long-range) interactions.
Employing the decomposition in Eq.~(\ref{eq:decomposition_2}) is thus useful when dealing with $\chi$ satisfying $\tilde{N}_{\rm int}\chi \gg N^{~}_{\rm int}$.
%\begin{equation}
%\tilde{H} = \sum_{p} \tilde{H}^{(p)}_{0} + \sum_{\ell=1}^{N^{~}_{\rm int}} \tilde{O}^{(p_\ell)}_{\ell} \tilde{O}^{(p'_\ell)}_{\ell}
%\label{eq:decomposition_2}
%\end{equation}
%with $p$ and $p_\ell \in \{ a,b,c,d \}$, where $\tilde{H}^{(p)}_{0}$ is a renormalized Hamiltonian in the subsystems specified by bonds $p$. Then, $\tilde{O}^{(p_\ell)}_{\ell}$ are renormalized operators in the subsystems specified by bonds $p_\ell$ to describe the interactions between four subsystems, and $N^{~}_{\rm int}$ is of the order of the number of interactions between the subsystems and scales with sum of surface area (volume) of the smaller of the two subsystems in volume for all patterns where two subsystems are extracted from the four subsystems when the system consists of short-range (long-range) interactions. 
%A computational cost of the order of $(4+2N^{~}_{\rm int})\chi^5$ is required for the operation of $\tilde{H}{\tilde \Psi}$ by use of the decomposition in Eq.~(\ref{eq:decomposition_2}).
%This prescription is useful when dealing with $\chi$ satisfying $(1+\tilde{N}_{\rm int})\chi > 2+N^{~}_{\rm int}$.

%Under conditions where the prescription is valid, the next 
Another computational hotspot with the cost of $O(\chi^6)$ is the full diagonalization of the reduced DM, whose dimension is of the order of $\chi^2$ in the TTN calculation.
Here, if we do not need to care about the degeneracy of singular values due to the symmetry of the target state, the partial SVD may effectively reduce the cost down to $O(\chi^5)$~\cite{Halko_2011}.
The randomized SVD may further reduce the cost to $O(\chi^4 \log\chi)$.

The computational costs in the TTN calculation discussed above are much larger than the corresponding ones in the DMRG since in the MPN employed in the DMRG, two of the four bonds entering the central area, two out of $\{ a,b,c,d \}$ in Fig.~\ref{fig:process_1}, have the dimension $\nu$, which is usually much smaller than $\chi$.
As a result, the cost for the diagonalization of $\tilde{H}$ represented in the decomposition in Eq.~(\ref{eq:decomposition_2}) is $O(N^{~}_{\rm int} \nu^2\chi^3)$ in the DMRG, being much smaller than $O(N^{~}_{\rm int}\chi^5)$ in the TTN algorithm.
Similarly, the cost for the full diagonalization of the reduced DM in the DMRG is $O(\nu^3\chi^3)$, that is significantly small compared with $O(\chi^6)$ in the TTN algorithm.

\section{Coordinates of auxiliary bonds in Fig.\ \ref{fig:Hier-EE}}\label{App:coordinate}

In Fig.\ \ref{fig:Hier-EE}, we have introduced the coordinates to indicate the position of the auxiliary bonds.
The horizontal coordinate $x$ introduced for both the optimized TTN and MPN are defined as follows.
We first assign the site index $i$ to the horizontal coordinate of bare spins.
The position of an isometry is then given by the average of the horizontal coordinates of bare spins or the isometries connected with the incoming bonds.
The position $x$ of the auxiliary bonds is finally defined by the average of the coordinates of the connected isometries. 

For the case of the optimized TTN, moreover, we have introduced the height coordinate $y$, which is defined as the distance of the bond from the nearest bare spin;
We assign $y=0$ for the bonds directly connected to a bare spin, and $y$ increases by one as the position of the bond moves away from the bare spin.

\begin{figure}[b]
\begin{center}
\includegraphics[width = 75 mm]{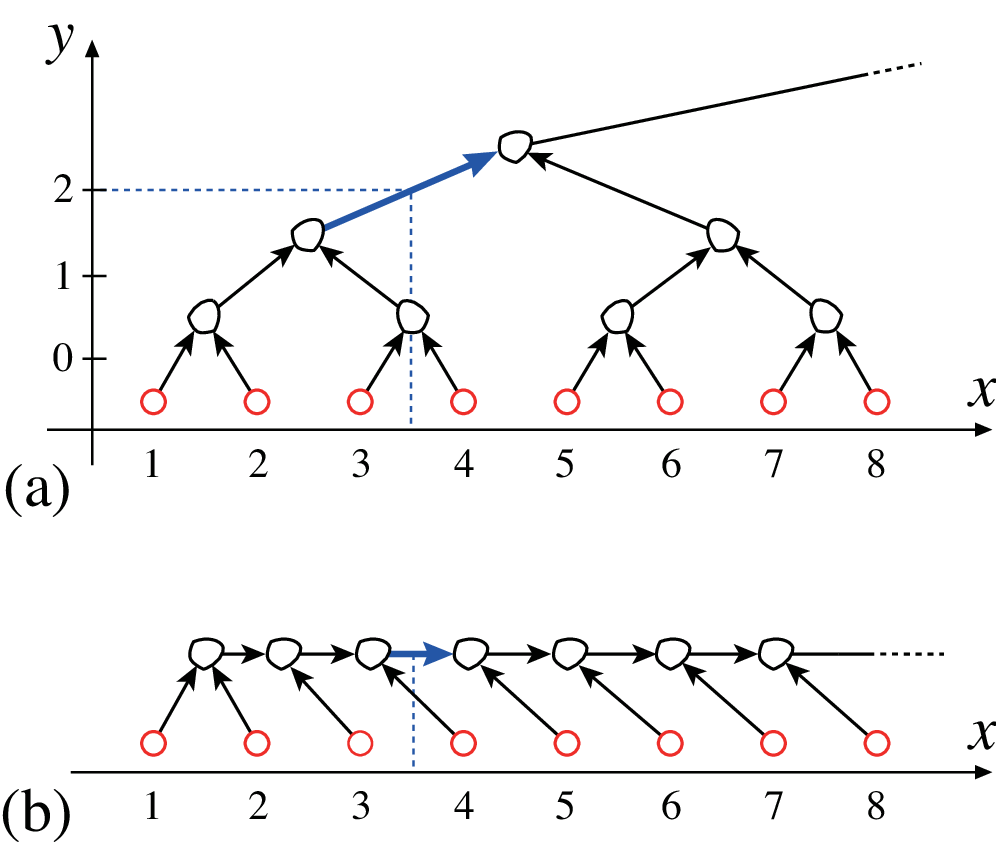}
\caption{
Schematic pictures for the coordinates indicating the positions of the auxiliary bonds for (a) the optimized TTN and (b) the MPN. 
In (a), the coordinates for the bond denoted by a thich arrow (blue) are $(x,y) = (3.5, 2)$.
In (b), the coordinate for the bond is $x=(3.125+4.0625)/2=3.59375$.
}
\label{fig:coordinate}
\end{center}
\end{figure}

Figure\ \ref{fig:coordinate} presents the schematic pictures to illustrate the coordinates in the optimized TTN and the MPN.
We note that the horizontal coordinate $x$ for the isometries and auxiliary bonds in the MPN is shifted towards the open edges of the system compared to the diagram shown in Fig.\ \ref{fig:Hier-trans} (a), where we do not use the coordinate $x$ for readability.

%\bibliography{TTNopt_Ref}

\begin{thebibliography}{77}
\bibitem [{not()}]{note_MPN}%
  \BibitemOpen
  \bibinfo {note} {We use a TTN as a term indicating the amplitudes of a
  wavefunction represented in the form of a contraction of tensors on a tree
  network. Meanwhile, the ket state $|\Psi\rangle$ with a TTN is termed TTN
  state. Given a MPS, similarly, the MPN represents the amplitudes of the
  corresponding wavefunction.}\BibitemShut {Stop}%
\end{thebibliography}


\end{document}